\documentstyle[aps,prl,twocolumn,psfig,floats]{revtex}     

\begin{document}
\preprint{\vbox{\hbox{{\tt hep-ph/0108115}\\ August 2001}}}
\draft
\wideabs{
\title{Shape versus Volume:  Making Large Flat Extra Dimensions Invisible}
\author{Keith R. Dienes}
\address{Department of Physics, University of Arizona, Tucson, AZ  85721 USA}
\address{E-mail address:  ~{\tt dienes@physics.arizona.edu}}
\date{August 13, 2001}
\maketitle
\begin{abstract}
     Much recent attention has focused on theories with 
     large extra compactified dimensions.
     However, while the phenomenological implications of the volume moduli 
     associated with such compactifications are well understood, 
     relatively little attention has been devoted to the 
     shape moduli.  In this paper, we show that
     the shape moduli have a dramatic effect on the corresponding Kaluza-Klein  
     spectra:  they change the mass gap, induce level crossings, and can even
     be used to interpolate between theories with different
     numbers of compactified dimensions.  Furthermore, we show that 
     in certain cases it is possible to maintain the 
     ratio between the higher-dimensional and four-dimensional Planck
     scales while simultaneously increasing the Kaluza-Klein graviton mass
     gap by an arbitrarily large factor.
     This mechanism can therefore be used to alleviate
     (or perhaps even eliminate) many of the experimental bounds on 
     theories with large extra spacetime dimensions.  
\end{abstract}
          }

\newcommand{\newc}{\newcommand}
\newc{\gsim}{\lower.7ex\hbox{$\;\stackrel{\textstyle>}{\sim}\;$}}
\newc{\lsim}{\lower.7ex\hbox{$\;\stackrel{\textstyle<}{\sim}\;$}}

\def\beq{\begin{equation}}
\def\eeq{\end{equation}}
\def\beqn{\begin{eqnarray}}
\def\eeqn{\end{eqnarray}}
\def\half{{\textstyle{1\over 2}}}
\def\ie{{\it i.e.}\/}
\def\eg{{\it e.g.}\/}


\def\inbar{\,\vrule height1.5ex width.4pt depth0pt}
\def\IR{\relax{\rm I\kern-.18em R}}
 \font\cmss=cmss10 \font\cmsss=cmss10 at 7pt
\def\IQ{\relax{\rm I\kern-.18em Q}}
\def\IZ{\relax\ifmmode\mathchoice
 {\hbox{\cmss Z\kern-.4em Z}}{\hbox{\cmss Z\kern-.4em Z}}
 {\lower.9pt\hbox{\cmsss Z\kern-.4em Z}}
 {\lower1.2pt\hbox{\cmsss Z\kern-.4em Z}}\else{\cmss Z\kern-.4em Z}\fi}

\input epsf


\section{Introduction}

Over the past several years, there has been an explosion of interest
in theories with large extra spacetime dimensions.  Much of this 
interest stems from the realization that large extra dimensions
have the potential to lower the fundamental energy scales of physics,
such as the Planck scale~\cite{ADD}, the GUT scale~\cite{DDG}, and the 
string scale~\cite{string}.
Indeed, as is well understood, the degree to which these scales 
may be lowered depends on the volume of the compactified dimensions.  

However, compactification manifolds are generally described by 
shape moduli (so-called ``complex moduli'') as well as volume moduli
(so-called ``K\"ahler moduli'').  This distinction has phenomenological
relevance because the shape moduli also play a significant role in
determining the experimental bounds on such scenarios.  Unfortunately,
in most previous discussions of extra dimensions, relatively little attention  
has been paid to the implications of these moduli.

In this paper, we shall discuss the phenomenological implications of 
the shape moduli by focusing on the simple case of a flat, 
two-dimensional toroidal compactification.  In this case, the relevant shape
modulus corresponds to the relative angle $\theta$ between the two
directions of compactification.  As we shall demonstrate,
the corresponding Kaluza-Klein spectrum is strongly dependent on $\theta$,
and exhibits level-crossing as well as a changing mass gap as $\theta$ is
varied.  This indicates that shape moduli such as  $\theta$ should not be ignored
in phenomenological studies of large extra dimensions.  Moreover, we shall see
that such shape moduli even provide an interesting means of {\it interpolation}\/  
between theories with different numbers of extra spacetime dimensions.
Finally, we shall show that under certain circumstances,
it is possible to exploit shape moduli in order to increase
the Kaluza-Klein mass gap by an arbitrarily large factor;
this occurs even though the volume of compactification remains fixed.
This surprising observation can therefore be used to alleviate (and 
perhaps even eliminate) many of the bounds that currently 
constrain such theories with large extra dimensions.

\section{Compactification on a two-torus with shift angle:  ~Kaluza-Klein spectrum}

\begin{figure}[b]
\centerline{
      \epsfxsize 2.2 truein \epsfbox {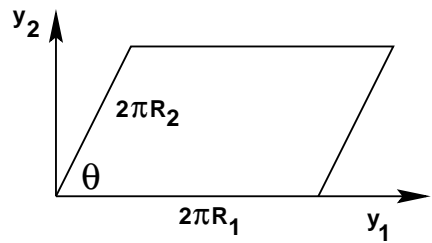}}
\caption{General two-dimensional torus with shift angle $\theta$.}
\label{torus}
\end{figure}

Since one-dimensional compactifications lack shape moduli, we begin
the discussion by considering compactification on a general two-torus,
as shown in Fig.~\ref{torus}.  Such a torus is specified by three
real parameters (the two radii $R_1,R_2$ of the torus as well as
the shift angle $\theta$),
and corresponds to identifying points which are related under 
the two coordinate transformations
\beqn
         && \cases{  y_1 ~\to~ y_1+2\pi R_1\cr
                     y_2 ~\to~ y_2\cr}  \nonumber\\
         && \cases{  y_1 ~\to~ y_1+2\pi R_2 \cos\,\theta\cr
                     y_2 ~\to~ y_2+2\pi R_2 \sin\,\theta~.\cr}
\label{torusdef}
\eeqn
Note that we are using orthogonal coordinates $y_i$ for the
extra dimensions;  likewise, since this is a toroidal compactification,
the metric remains {\it flat}\/ for all angles $\theta$.
As evident from Eq.~(\ref{torus}), the physical significance of the angle $\theta$ 
is that translations along the $R_2$ direction produce simultaneous
translations along the $R_1$ direction.  
Note that tori with different angles $\theta$ are 
topologically distinct (up to the modular transformations to be
discussed below).  

There are two  ``shape'' parameters for such a torus:  
the ratio $R_2/R_1$ and the angle $\theta$.
While most previous discussions of large extra dimensions have focused
on the volume of such tori (and even on the ratio $R_2/R_1$), they
have ignored the possibility of the shift $\theta$, essentially fixing
$\theta=\pi/2$.  Our goal, therefore,
is to understand the phenomenological implications of the angle $\theta$.

Given the torus identifications in Eq.~(\ref{torusdef}), it is straightforward
to determine the corresponding Kaluza-Klein spectrum. 
The Kaluza-Klein eigenfunctions for such a torus are given by
\beq
     \exp\left\lbrack i {n_1\over R_1} \left(y_1- {y_2\over \tan\theta}\right) 
                 ~+~ i{n_2\over R_2} {y_2\over \sin\theta} \right\rbrack~
\label{KKfuncts}
\eeq
where $n_i\in \IZ$.
Applying the (mass)$^2$ operator 
    $- (\partial^2 /\partial y_1^2 + \partial^2 /\partial y_2^2)$, 
we thus obtain the corresponding
Kaluza-Klein masses
\beq
    M_{n_1,n_2}^2 ~=~ {1\over \sin^2\theta} \left(
         {n_1^2\over R_1^2} +   
         {n_2^2\over R_2^2} - 2 {n_1n_2\over R_1R_2} \cos\theta\right)~.
\label{KKmasses}
\eeq
We see that while the Kaluza-Klein spectrum maintains its invariance 
under $(n_1,n_2)\to -(n_1,n_2)$,
it is no longer invariant under
$n_1\to -n_1$ or $n_2\to -n_2$ individually.  The spectrum is, however,
invariant under either of these shifts and the simultaneous shift 
$\theta\to \pi-\theta$.
We can therefore restrict our attention to tori with angles in the range
$0<\theta\leq \pi/2$ without loss of generality.

\begin{figure}[ht]
\centerline{
      \epsfxsize 3.6 truein \epsfbox {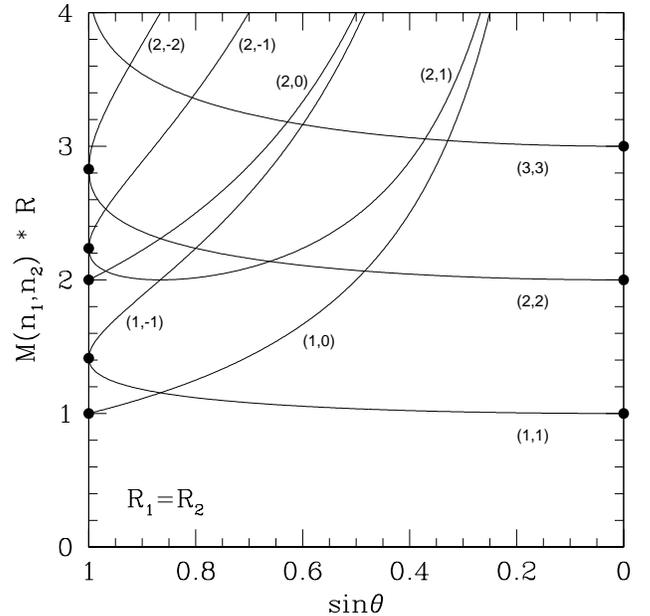}}
\caption{The lowest-lying Kaluza-Klein spectrum 
        as a function of the shape parameter $\theta$ 
        for $R_1=R_2\equiv R$.  
        In general, for $R_1=R_2$, each line labelled by $(n_1,n_2)$
        represents the four-fold degeneracy of states 
        $\lbrace \pm(n_1,n_2), \pm(n_2,n_1)\rbrace$ for all $\theta$;  
        this degeneracy is only two-fold if $n_1=n_2$ or if either $n_1$ or $n_2$
        vanishes.
     }
\label{equalRfig}
\end{figure}

It is clear from Eq.~(\ref{KKmasses}) that the Kaluza-Klein masses
depend on $\theta$ in a non-trivial, level-dependent way.  
In order to deduce the physics behind Eq.~(\ref{KKmasses}), let
us first examine the case with $R_1=R_2\equiv R$.
We then find the results shown in Fig.~\ref{equalRfig}.
As guaranteed by Eq.~(\ref{KKmasses}), the ground state remains 
massless for all $\theta$.  However, as $\theta$ is varied,
we see from Fig.~\ref{equalRfig}
that the excited Kaluza-Klein spectrum exhibits dramatic changes,  
with many light states becoming heavy and several heavier states 
becoming light.

Interestingly, as a result of this level-crossing,
the identity of the lowest excited state is itself a function of 
$\theta$, with the $(\pm 1,0)$ and $(0,\pm 1)$ states
(four states total)
serving as the lowest excitations for $\pi/3\leq \theta \leq \pi/2$,
and the $\pm(1,1)$ states (two states total) filling this role 
for $\theta\leq \pi/3$.
In general, we observe that both the {\it mass gap}\/ $\mu$ (defined
as the splitting
between the ground state and the first excited states)
and the {\it degeneracy}\/ $\alpha$ of the first excited states 
are functions of the shape parameter $\theta$.  

This is particularly important in the case of Kaluza-Klein gravitons.
In general, the presence of Kaluza-Klein gravitons induces
deviations from Newtonian gravity, with the corresponding 
gravitational potential
taking the form~\cite{ADD,kehagias}
\beq
       V(r) ~=~ - G_4 {m_1 m_2\over r}\, ( 1+ \alpha e^{-\mu r} +...)
\label{newton}
\eeq
for $r\gg 1/\mu$.
Thus, in this simple case with $R_1=R_2$, 
we see that the expected deviations from non-Newtonian gravity 
drop by a factor of two when $\theta < \pi/3$ --- even though
the radii are held fixed.

It is also possible to understand the behavior of the Kaluza-Klein
spectrum as $\theta\to 0$.  
In this limit, the two cycles of the torus collapse onto each
other.  The resulting Kaluza-Klein spectrum therefore depends
on whether the periodicities of the two cycles are commensurate.
In the case with $R_1=R_2\equiv R$, the two cycles are commensurate, and
the torus identifications in Eq.~(\ref{torusdef}) collapse to become the 
single identification corresponding to a circle of radius $R$.  
This behavior is apparent in Fig.~\ref{equalRfig}:  the only Kaluza-Klein 
states which remain light as $\theta\to 0$ are those which effectively 
reproduce a one-dimensional circle-compactification with radius $R$.
Thus, we see that the shape parameter $\theta$ allows us to 
smoothly {\it interpolate}\/ between compactifications of different
numbers of spacetime dimensions.  Note, in particular, that this method 
of interpolation is physically different from the standard method 
of interpolation in which a single radius is taken to infinity.

This interpolation behavior as $\theta\to 0$
is completely general, and arises for all rational 
values of $R_2/R_1$.  
In the limit $\theta\sim\epsilon\ll 1$, Eq.~(\ref{KKmasses}) becomes
\beqn
      M_{n_1,n_2}^2 &\approx& {1\over \epsilon^2}
                 \left( {n_1\over R_1}-{n_2\over R_2}\right)^2 ~+~\nonumber\\
        && +~  {1\over 3} \left( {n_1^2\over R_1^2} + {n_2^2\over R_2^2} +  
          {n_1 n_2\over R_1 R_2}\right)  ~+~ {\cal O}(\epsilon^2) ~.
\label{limitcase}
\eeqn
We thus see that $M_{n_1,n_2}\to \infty$ as $\epsilon\to 0$ for all 
$(n_1,n_2)$ unless 
$R_2/R_1$ is a rational number.  In these cases, 
we may represent $R_2/R_1= p/q$ where $p$ and $q$ are relatively
prime.  We then find that the radius of the resulting circle-compactification
as $\theta\to 0$
is given by $R\equiv R_1/q = R_2/p$, with the torus modes 
$(n_1,n_2)=k(q,p)$ evolving to become the circle modes $M_k=k/R$
and all others becoming infinitely massive.
Note, in particular, that when $p,q\not= 1$, the radius $R$ of the resulting
circle compactification is generally {\it smaller}\/ than either $R_1$ or
$R_2$.  This implies that the corresponding circle Kaluza-Klein states 
are {\it heavier}\/ than the initial  
torus Kaluza-Klein states with which we started.

\begin{figure}[h]
\centerline{
      \epsfxsize 3.6 truein \epsfbox {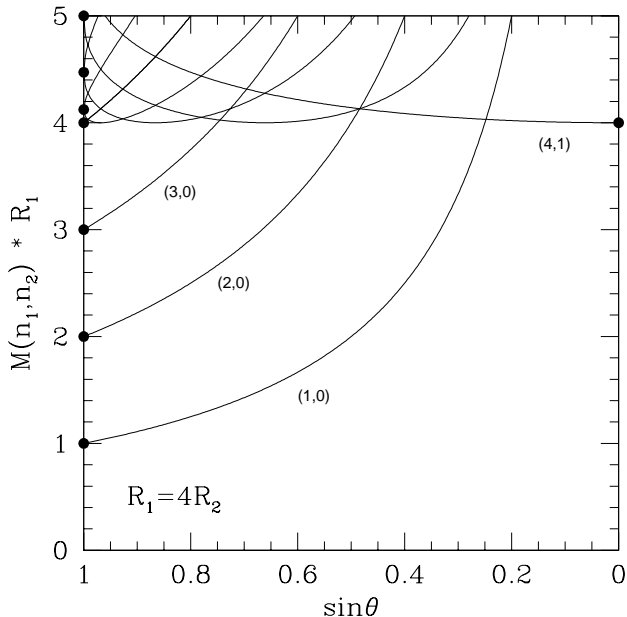}}
\caption{The Kaluza-Klein spectrum as a function of the shape parameter
       $\theta$ for $R_1=4 R_2$.  Each state $(n_1,n_2)$ is two-fold 
         degenerate with $-(n_1,n_2)$. 
     }
\label{onequarter}
\end{figure}

This behavior when $R_2/R_1$ is rational
is illustrated in Fig.~\ref{onequarter}, where
we have taken $R_2/R_1=1/4$.
In this case, the radius of the resulting circle compactification
is $R_1/4$.  The mass gap therefore becomes four times as large as $\theta\to 0$
as it was at $\theta=\pi/2$, leading to an {\it exponential}\/ 
suppression of the deviations from non-Newtonian gravity.

It is interesting to explore the case when $R_2/R_1$ is {\it not}\/
a rational number.  Indeed, unless there is some dynamics that fixes
the radius moduli to have a rational ratio, this will be the generic
situation.  In such cases, {\it all}\/ excited Kaluza-Klein  
states become infinitely massive as $\theta\to 0$;  essentially
the radius of the resulting circle-compactification is zero.
This divergence of the Kaluza-Klein masses is ultimately a reflection
of the incommensurate nature of the two torus periodicities, an
incompatibility which grows increasingly severe as $\theta\to 0$.
Note that the actual limit as $\theta\to 0$ is a singular one,
corresponding to a degenerate compactification manifold.
However, our point is that when $R_2/R_1$ is irrational,
we can always make our excited
Kaluza-Klein states arbitrarily heavy by choosing a sufficiently 
small value for $\theta$.  Thus, for all intents and purposes, 
there always exists a (small, non-zero) value of $\theta$ for which 
we can make our extra dimensions
truly ``invisible'' with respect to laboratory or observational
constraints that rely on the presence of light Kaluza-Klein states.

\section{Shape versus Volume}

Thus far, we have shown that when $R_2/R_1$ is irrational,
our excited Kaluza-Klein 
states become arbitrarily heavy as $\theta\to 0$. 
As such, these
extra dimensions become ``invisible'', even though  
the radii $R_1,R_2$ are held fixed.
However, even though the radii are held
fixed, the {\it volume}\/ of the extra dimensions is falling
to zero.
Indeed, when $R_2/R_1$ is irrational,
the compactification volume falls like $\sin\theta$ while the excited Kaluza-Klein masses
diverge as $1/\sin\theta$.
To what extent, then, 
can the compactification volume remain ``large''
while the corresponding Kaluza-Klein states become heavy?
Indeed, what is the role of the shape moduli when the volume is held fixed? 
As we shall see, this issue is surprisingly subtle.

To address these issues, we now study the behavior of the
Kaluza-Klein masses when the {\it volume}\/ of the compactification manifold
is held fixed.
Towards this end, 
let us reparametrize the three torus 
moduli $(R_1,R_2,\theta)$ 
in terms of a single real volume modulus $V$ 
and a {\it complex}\/ shape modulus $\tau$:
\beq
    V \equiv 4\pi^2 R_1 R_2 \sin\theta~,~~~~~~~
    \tau \equiv {R_2\over R_1}\,e^{i\theta}~.
\label{moduli}
\eeq
We shall also define $\tau_1\equiv {\rm Re}\,\tau$ and
$\tau_2\equiv {\rm Im}\,\tau$.
Using these definitions, we can express
$(R_1,R_2,\theta)$ in terms of $(V,\tau)$ via
\beqn
        && \cos\theta = \tau_1/|\tau| ~,~~~~ \sin\theta = \tau_2/|\tau|~,\nonumber\\ 
        && R_1^2 =  {1\over 4\pi^2 \tau_2}V~,~~~~
            R_2^2 =  {|\tau|^2 \over 4\pi^2 \tau_2}V~.
\eeqn
The periodicities in Eq.~(\ref{torusdef})
then take the form
\beq
       z\to z+ \sqrt{{V\over \tau_2}} ~,~~~~~~
       z\to z+ \sqrt{{V\over \tau_2}}\,\tau ~
\eeq
where $z\equiv y_1+iy_2$,
and the Kaluza-Klein wavefunctions in Eq.~(\ref{KKfuncts})
take the form
\beq
       \exp \left\lbrace 
         {2\pi i\over \sqrt{V\tau_2}} \,{\rm Im}\,  
              \lbrack \overline{z} \,(n_1\tau-n_2)\rbrack \right\rbrace~.
\eeq
Operating with the (mass)$^2$ operator 
$-4 \partial^2/(\partial z \partial \overline{z})$
then yields the Kaluza-Klein masses 
\beqn
       M_{n_1,n_2}^2 &=& {4\pi^2 \over V} {1\over \tau_2} 
          \bigl| n_1\tau - n_2 \bigr|^2~\nonumber\\
         &=& {4\pi^2 \over V} {1\over \tau_2} 
      \left\lbrack  (n_1\tau_1 - n_2)^2 + n_1^2 \tau_2^2 \right\rbrack~.
\label{KKmassestwo}
\eeqn
Note that although Eq.~(\ref{KKmassestwo}) is merely a rewriting 
of Eq.~(\ref{KKmasses}), we have now explicitly separated the
effects of the volume modulus $V$ from those of the shape modulus $\tau$.
Writing the remaining shape factors in terms of the original 
parameters $(R_1,R_2,\theta)$, we thus obtain
\beq
   \left({ V\over 4\pi^2 }\right) M_{n_1,n_2}^2 ~=~ 
    {1\over \sin\theta} \left\lbrack
      n_1^2 {R_2\over R_1} + n_2^2 {R_1\over R_2} - 2 n_1n_2 \cos\theta\right\rbrack~.
\label{aneweq}
\eeq

\begin{figure}
\centerline{ \epsfxsize 3.6 truein \epsfbox {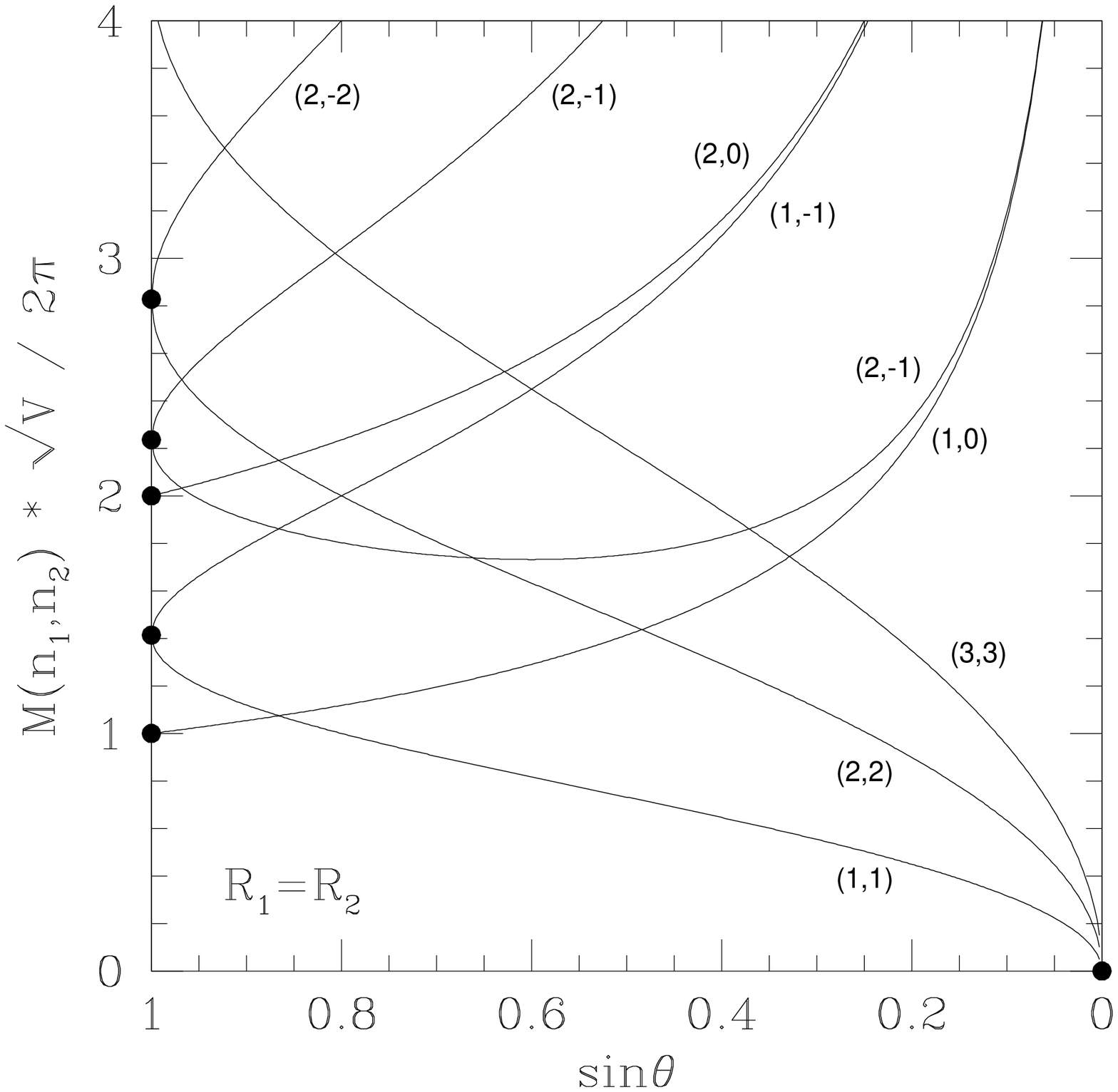}}
\centerline{ \epsfxsize 3.6 truein \epsfbox {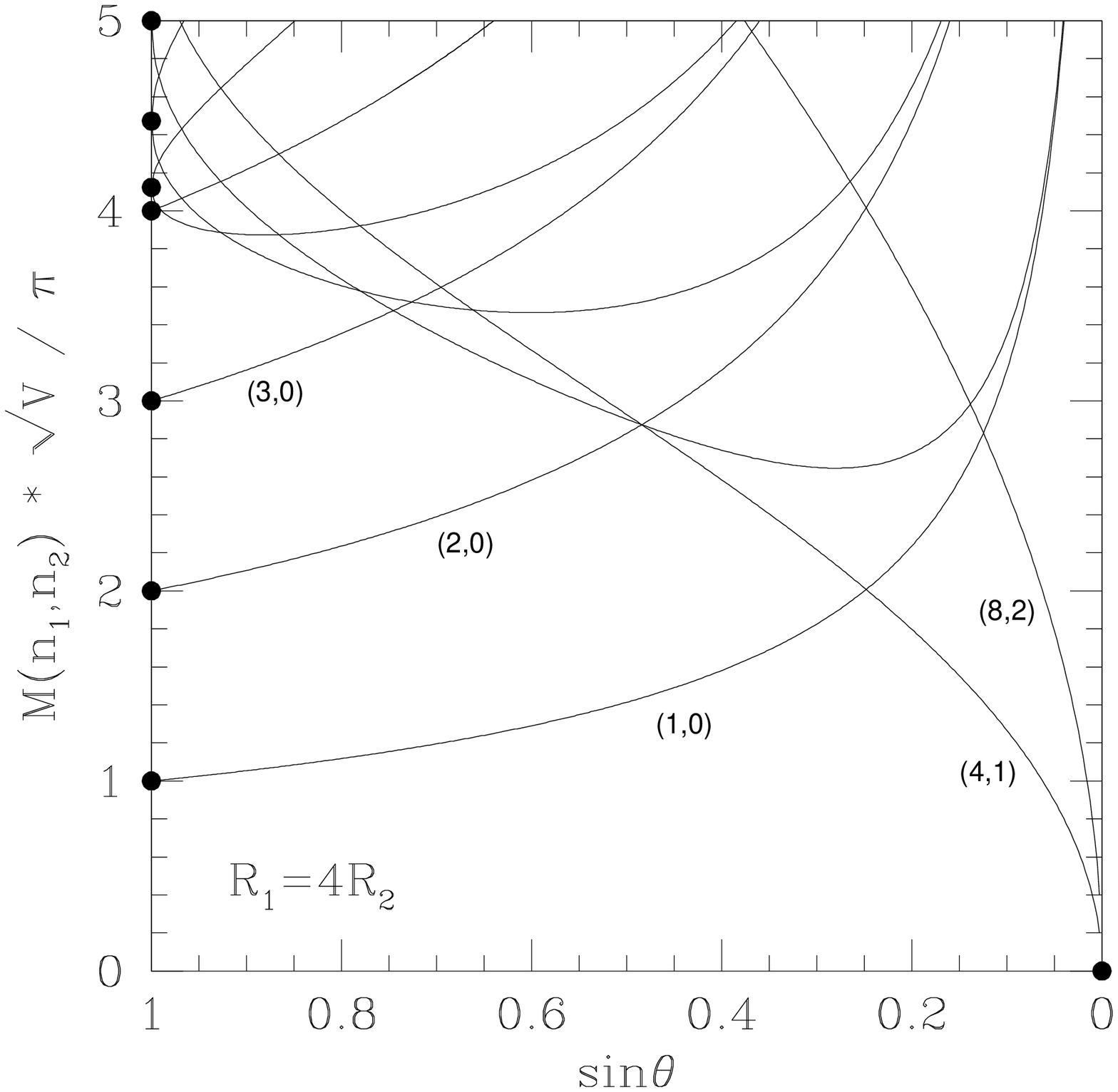}}
\caption{The Kaluza-Klein mass spectrum as a function of the shape parameter
     $\theta$ for $R_1= R_2$ (upper plot) and $R_1=4 R_2$ (lower plot).  
    In each case, the compactification volume is held fixed.}
\label{newfig}
\end{figure}

Using this result, we can now consider the effects of the 
shape parameter $\theta$ when the compactification {\it volume}\/ is 
held fixed.
In Fig.~\ref{newfig}, we plot the Kaluza-Klein masses for the 
$R_1=R_2$ and $R_1=4 R_2$ cases 
considered earlier in Figs.~\ref{equalRfig}~and~\ref{onequarter}.
Note that in order to keep the volume fixed as $\theta\to 0$, the radii 
are now forced to grow increasingly large (even though their ratio is held fixed).
This increase in the radii therefore provides an extra 
tendency towards {\it lowering}\/ the Kaluza-Klein masses, as
can be seen by comparing the masses
plotted in Fig.~\ref{newfig} to those plotted  
in Figs.~\ref{equalRfig}~and~\ref{onequarter}. 

Despite this tendency towards smaller Kaluza-Klein masses, 
we see that the shape moduli can help to render
the large extra dimensions effectively invisible even when the compactification
volume is held fixed.  For example, in the case with $R_1=4R_2$, 
we see that the mass gap increases by a factor of two
near $\sin\theta\approx 1/4$.
This occurs even though the ratio $R_2/R_1$ and the compactification volume
are both being held fixed relative to their values at $\theta=\pi/2$.
Indeed, these suppressions become even more pronounced 
for scenarios with $R_2/R_1=p/q$ with larger values of $(p,q)$.
We stress that these features are possible only because of the
introduction of non-trivial shape moduli. 

Nevertheless, the utility of this mechanism is ultimately limited 
when $R_2/R_1$ is rational
because of the appearance of an infinite tower of Kaluza-Klein states
which become massless in the $\theta\to 0$ limit.
Of course, these states are nothing but the circle-compactification
states whose masses previously remained finite
when the radii were held fixed
in Figs.~\ref{equalRfig}~and~\ref{onequarter}. 
As discussed above, these states now become massless when the
volume is held fixed because the radii must now 
compensate by becoming infinitely large as $\theta\to 0$.

Given this, it is natural to wonder whether we may avoid the appearance
of these dangerous massless states in the $\theta\to 0$ limit by considering
the case when $|\tau|\equiv R_2/R_1$ is {\it irrational}\/.
It is immediately clear that there are two opposite forces at play 
in such cases.
First, there is the tendency towards masslessness
which arises because of the expanding radii as $\theta\to 0$.
However, this must compete against the opposite tendency, driven by the
irrationality of the ratio $R_2/R_1$, which
pushes the masses of the resulting ``circle-compactified'' 
states to infinity (or equivalently pushes the radius of the resulting 
``circle-compactification'' to zero).

In order to determine the net effect on the Kaluza-Klein mass spectrum,
let us return to Eq.~(\ref{aneweq}) and
consider the $\theta\sim\epsilon \ll 1$ limit:
\beqn
    && \left( {V\over 4\pi^2 }\right) M_{n_1,n_2}^2 ~=~  
             {(n_2-n_1 |\tau|)^2 \over |\tau| \epsilon}+\nonumber\\
     && ~~~~~~~~~~~~+ \left( {n_2^2 + 4 n_1 n_2 |\tau| + n_1^2 |\tau|^2 \over 6 |\tau|}
             \right) \epsilon +  {\cal O}(\epsilon^3)
\label{limitcasetwo}
\eeqn
where $|\tau|\equiv R_2/R_1$ is irrational.
This is the fixed-volume analogue of Eq.~(\ref{limitcase}).
Given this expression, we can immediately see the two tendencies at work.
The first term on the right side of Eq.~(\ref{limitcasetwo})  
generally diverges because $n_2-n_1|\tau|$ never vanishes exactly.
Thus, the general Kaluza-Klein state becomes infinitely heavy
as $\theta\to 0$. 
However, for any fixed chosen value of $\epsilon$, we can always find
special states $(n_1,n_2)$ for which this first term comes {\it arbitrarily close}\/ 
to cancelling;  this simply requires choosing sufficiently large values 
of $(n_1,n_2)$.  These special states with large $(n_1,n_2)$ are potentially massless. 
On the other hand, choosing such large values of $(n_1,n_2)$ drives the second
term in Eq.~(\ref{limitcasetwo}) to larger and larger values.  [Note that the third and higher terms are
always suppressed relative to the second term in the $\epsilon\to 0$ limit,
even as $(n_1,n_2)$ grow large.]  
Thus, because of the conflict between these two terms,
it is not readily obvious whether these special, potentially massless states
actually become massless in the $\theta\to 0$ limit. 

The outcome of this competition between the first two terms in
the Kaluza-Klein mass formula in Eq.~(\ref{limitcasetwo}) rests
on the efficiency with which $n_2 - n_1 |\tau|$ can be made to approach zero 
for integer $(n_1,n_2)$, as a function of $n_2$, 
given an arbitrary irrational number $|\tau|$.
Let us parametrize this efficiency in the form
\beq
           (n_2 - n_1 |\tau|)^2  \sim {A^2\over n_1^{2(1+\gamma)}}  ~~~ 
                    {\rm as}~ (n_1,n_2)\to \infty~
\label{numbtheory}
\eeq
for some constants $A$ and $\gamma$.
We shall see shortly that this is indeed the most relevant 
parametrization for this asymptotic behavior.
Thus, for these potentially massless states, Eq.~(\ref{limitcasetwo}) 
becomes
\beq
    \left( {V\over 4\pi^2 }\right) M_{n_1,n_2}^2 ~=~  
             { A \over n_1^\gamma } \left( y + {1\over y}\right) +...
\label{yeq}
\eeq
where $y\equiv n_1^{2+\gamma}|\tau| \epsilon/A$.

Fortunately, $y+y^{-1}$ is bounded from below for all values of $y$.
The issue therefore boils down to a simple number-theoretic question:
what is the value of $\gamma$?
Clearly, if $\gamma>0$, we see from Eq.~(\ref{yeq}) that the
lightest Kaluza-Klein states for irrational $|\tau|$ 
become massless as $\theta\to 0$.  By contrast, if $\gamma\leq 0$, 
then irrationality succeeds in preventing the appearance
of massless states as $\theta\to 0$.
 
It turns out that the value of $\gamma$ has been extensively
investigated in the mathematical literature. 
Indeed, this is nothing but the ancient problem of Diophantine 
approximation, with $2(1+\gamma)$ traditionally known as 
the ``irrationality measure'' or as the ``Liouville-Roth constant''.
The results are as follows~\cite{textbook}.
According to an 1842 theorem by Dirichlet,
for all irrational numbers $|\tau|$
it is possible to find an infinite number of integer
pairs $(n_1,n_2)$ such that Eq.~(\ref{numbtheory}) holds with
$\gamma \geq 0$.
However, for the case of {\it algebraic}\/ irrational
numbers $|\tau|$ (defined as irrational numbers 
which can be realized as solutions of non-zero polynomials with
integral coefficients), a stronger 1955 theorem due to Roth~\cite{textbook,Roth}
states that $\gamma \leq 0$.
(No such stronger theorem has yet been proven for non-algebraic
irrational numbers.)
Combining these two results, 
we conclude that $\gamma=0$ for the case of algebraic 
irrational numbers.  

This result indicates that {\it irrationality succeeds in preventing
the appearance of massless Kaluza-Klein states as $\theta\to 0$, even
if we hold the compactification volume fixed and 
the radii become infinitely large}\/.  
This result is
rigorous for algebraic irrational values of $|\tau|\equiv R_2/R_1$, and is
likely (though unproven) to hold for certain non-algebraic (transcendental)
irrational values as well.
Thus, for algebraic irrational numbers $|\tau|$, we see that these 
dangerous Kaluza-Klein states all have masses
which are bounded from below:
\beq
    \left( {V\over 4\pi^2 }\right) M_{n_1,n_2}^2  ~\geq~  {2 A}~.
\label{lowerbound}
\eeq
In other words, the {\it mass gap}\/ as $\theta\to 0$ is bounded from below according
to Eq.~(\ref{lowerbound}).
Moreover, for any given value of $\theta\sim \epsilon\ll 1$,
the Kaluza-Klein state which comes closest to saturating this bound 
is simply the state for which $y\approx 1$ in Eq.~(\ref{yeq}).
This is the state for which $n_1^2\approx A/(|\tau|\epsilon)$.

In order to measure the importance of the shape modulus $\theta$,
let us compare $\mu'$,
the mass gap in the $\theta\to 0$ limit,
with $\mu$, the original mass gap at $\theta=\pi/2$.
According to Eq.~(\ref{aneweq}),
the original (mass)$^2$ gap at $\theta=\pi/2$ is given
by either $|\tau|$ (if $|\tau|\leq 1$) or $1/|\tau|$ (if $|\tau|\geq 1)$.
Let us henceforth assume that $|\tau|\geq 1$ without loss of generality.
Thus, in general, the (mass)$^2$ gap as $\theta\to 0$ is greater than
the original (mass)$^2$ gap at $\theta=\pi/2$ by a factor
\beq
      \left({\mu'\over \mu}\right)^2 ~=~ 2 A |\tau|~.
\label{gapratio}
\eeq
Thus, if $2A|\tau| >1$, we have an {\it exponential}\/ suppression
of the effects of the Kaluza-Klein states as $\theta\to 0$.

\begin{figure}
\centerline{ \epsfxsize 3.6 truein \epsfbox {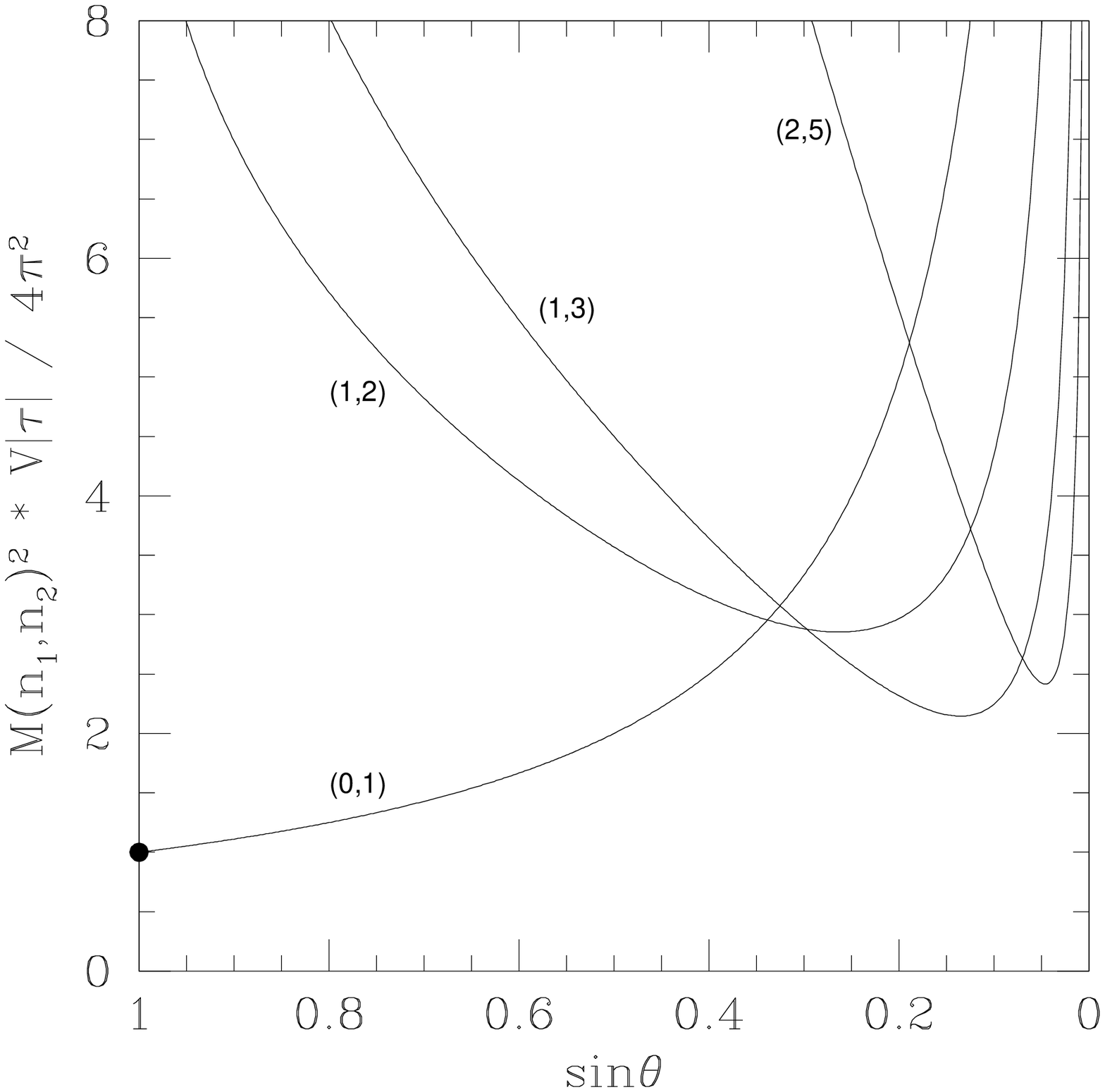}}
\centerline{ \epsfxsize 3.6 truein \epsfbox {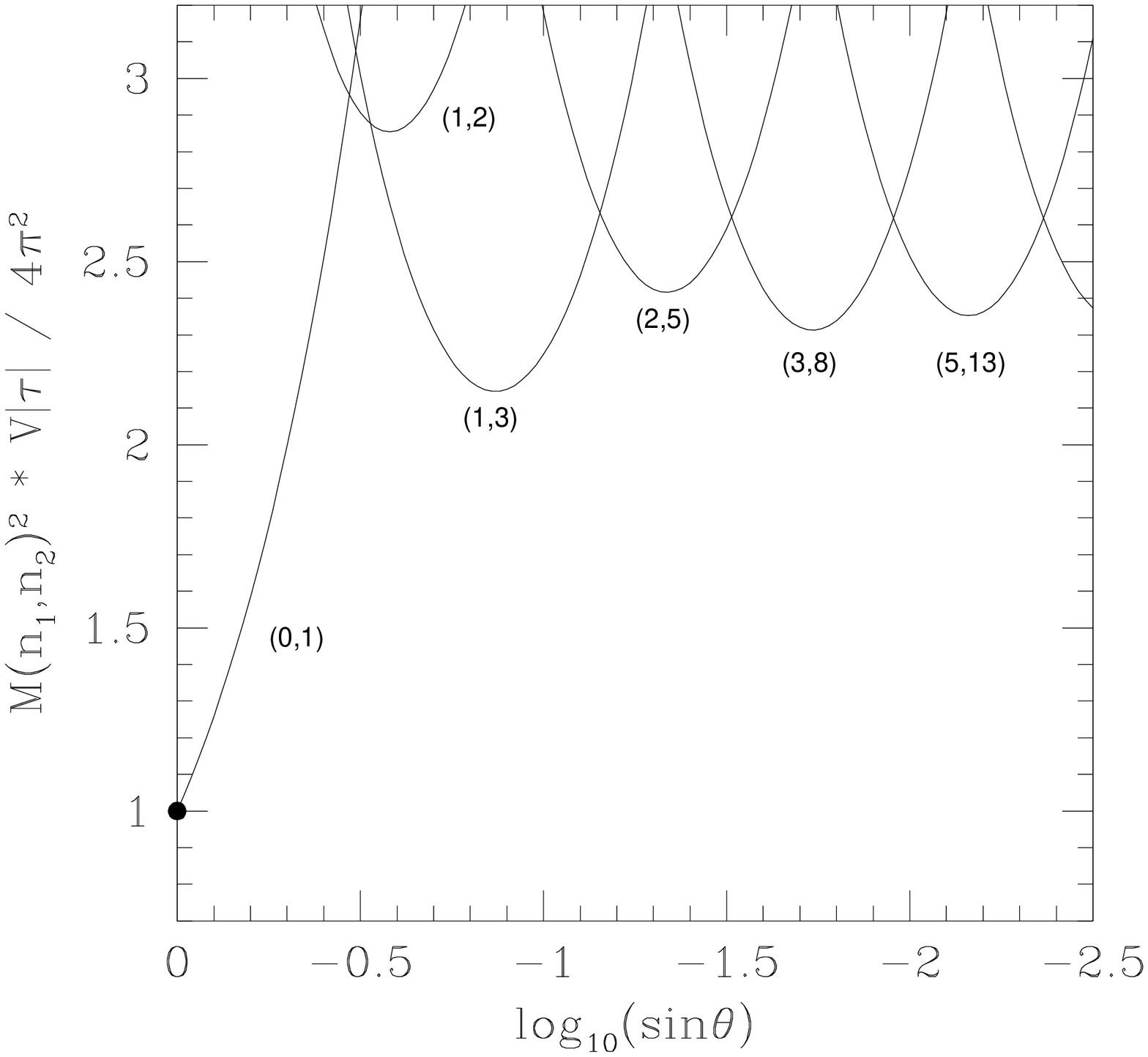}}
\caption{The Kaluza-Klein mass spectrum with $|\tau|\equiv R_2/R_1= \protect\half (3+\sqrt{5})$.
   In the upper plot~(a), we show the behavior of the $(0,1)$ 
    state as well as the behavior of several other low-lying Kaluza-Klein states
    which initially tend towards masslessness as $\theta\to 0$.  
   In the lower plot~(b), we illustrate the behavior of 
   these states as $\theta\to 0$ by plotting $\sin\theta$ on 
    a {\it logarithmic}\/ scale.
   In both plots, we have renormalized the overall Kaluza-Klein spectrum 
   so that the mass gap at $\theta=\pi/2$ is set to $1$.
   Note that the Kaluza-Klein mass gap is {\it larger}\/ as $\theta\to 0$
   than it is at $\theta=\pi/2$,
   even though the compactification volume $V$ and the radius ratio 
       $|\tau|\equiv R_2/R_1$
   are held fixed.} 
\label{logplot}
\end{figure}

Let us give an explicit example by considering a compactification
with $|\tau|\equiv R_2/R_1 = \half (3+\sqrt{5}) \approx 2.618$.
(We shall see later that this example is
motivated on both mathematical and physical grounds.) 
Our goal is to understand the corresponding spectrum of Kaluza-Klein states as 
we take $\theta\to 0$ while
holding $|\tau|$ and the compactification volume fixed.
As $\theta$ decreases, 
the above arguments indicate that the
majority of Kaluza-Klein states become increasingly heavy.
Indeed, the only states $(n_1,n_2)$ which have a tendency to become light
are those which are approximately ``circle-compactified'', satisfying
$n_2\approx|\tau| n_1$.  In this example with $|\tau| \approx 2.618$,
such low-lying states include $(1,2)$, $(1,3)$, $(2,5)$, {\it etc}\/.
The behavior of these states is illustrated in Fig.~\ref{logplot}(a).

Once $\theta$ becomes sufficiently small, however, the lightest 
(and hence most dangerous) states 
are ultimately those which lie on the leading 
$\gamma=0$ trajectory in Eq.~(\ref{numbtheory}). 
Which states are these?
While methods exist~\cite{textbook} for determining 
these leading states unambiguously for any value of $|\tau|$, 
in this case with $|\tau| = \half (3+\sqrt{5})$
it turns out that these leading states  
can easily be determined from the famous Fibonacci sequence
of integers $f_k =\lbrace 1,2,3,5,8,13,...\rbrace$
defined by the recursion relation $f_k=   f_{k-1}+f_{k-2}$ 
with $f_1=1$ and $f_2=2$.  
It is well known that the ratio of successive integers in this series
rapidly approaches the ``golden mean'' $g\equiv \half (1+\sqrt{5})$
as $k\to \infty$.
Thus, since $|\tau|=1+g$ in our example,
the lightest Kaluza-Klein states as $\theta\to 0$ are simply
the states $(n_1,n_2)=(f_{k},f_{k+2})$ for increasingly large values of $k$.
Note, in particular, that $f_{k+2}/f_{k} \to |\tau| = 1+g$ as $k\to \infty$,
as desired. 

It is straightforward to verify that this set of Kaluza-Klein
states satisfies Eq.~(\ref{numbtheory}) with $\gamma=0$.
This verifies that these states converge at the maximum possible rate ---
\ie, that these are indeed the lightest states as $\theta\to 0$.
However, since $\gamma=0$,
the masses of these states are bounded from below,
in accordance with Eq.~(\ref{lowerbound}). 
This behavior is illustrated in Fig.~\ref{logplot}(b).
Moreover, for each different value of $\theta$,
we see from Fig.~\ref{logplot}(b) that a different excited Kaluza-Klein 
state in the Fibonacci series has an enhanced tendency to become massless.
This is precisely the behavior discussed below Eq.~(\ref{lowerbound}).

It turns out that $A=1/\sqrt{5}$ in this example.
Since $|\tau| = 1+g = \half (3+\sqrt{5})$, 
the asymptotic mass gap ratio in this example is given by
$(\mu'/ \mu)^2  =  1 + 3/\sqrt{5}\approx 2.34$.   
This behavior is also shown in Fig.~\ref{logplot}(b),
where we have renormalized the overall Kaluza-Klein spectrum so that
the mass gap at $\theta=\pi/2$ is set to $1$.
Although not large in this example, this mass gap ratio
exceeds $1$ and thereby
leads to a relative {\it exponential}\/ suppression 
of the effects of the Kaluza-Klein states.
Specifically, recall from Eq.~(\ref{newton})
that in the case of Kaluza-Klein gravitons,
the corresponding deviations from Newtonian gravity 
are given by $\Delta V(r_\ast)  \sim \alpha e^{-\mu r_\ast}/r_\ast$
where $\alpha$ is the degeneracy of the first-excited Kaluza-Klein level,
$\mu$ is the mass gap, and 
$r_\ast\gg \mu^{-1}$ is the distance scale associated with such measurements.
A change in the mass gap from $\mu$ to $\mu'$ thereby suppresses these deviations
by an exponential factor 
\beq
      {\Delta V'\over \Delta V}  ~=~  {\alpha'\over \alpha} \,\exp\left\lbrack
           -\left( {\mu'\over \mu} -1\right) \mu r_\ast \right\rbrack~.
\label{suppressionfactor}
\eeq
Taking $\mu r_\ast\approx 10$ as a reference value,
we see that $\Delta V'/\Delta V\approx 5\times 10^{-3}$
for this example. 
This can therefore be a significant factor contributing to the 
invisibility of the extra dimensions, even though we have held 
both the compactification
volume and the radius ratio $|\tau|\equiv R_2/R_1$ fixed. 
         
Thus far, we have shown that irrationality succeeds in preventing 
the appearance of massless Kaluza-Klein states as $\theta\to 0$.
Indeed, we have seen that the Kaluza-Klein masses are 
always bounded from below, with the size of the mass gap 
as $\theta\to 0$ completely determined by the
value of the parameter $A$ in Eq.~(\ref{numbtheory}),
and the size of the mass gap ratio determined by the product $A|\tau|$
in Eq.~(\ref{gapratio}).
The next question, therefore, is to determine the sizes of $A$ and $A|\tau|$.
How large can these parameters become, and for what values of $|\tau|$ are
they maximized?

This issue has also been investigated in the mathematical literature,
with the following results.
It turns out that the set of all possible values of $A$ 
(the so-called ``Lagrange spectrum'') is generally discrete,
with only certain values allowed;  moreover, the set of all
algebraic irrational numbers $|\tau|$ may be sorted into equivalence
classes depending on their corresponding values of $A$.
According to a theorem by Hurwitz~\cite{textbook}, the Lagrange spectrum 
is bounded from above by the value $1/\sqrt{5}$.
Thus, our previous example already reaches the maximum possible value of $A$.  
However, the set of $|\tau|$ corresponding to each value of $A$
is countably infinite, and includes values of $|\tau|$ 
with ever-increasing magnitudes.
Thus, we see that although $A \leq 1/\sqrt{5}$, the value of the mass
gap ratio $2A|\tau|$ can be chosen to be {\it arbitrarily large}\/.

Once again, let us give an example.
Rather than consider $|\tau|= 1+g$ as in our previous example,
let us instead consider a more general compactification with the value 
$|\tau|=m + g$ where $m\in {\IZ}^+$. 
Such values of $|\tau|$ are all in the 
same equivalence class as the golden mean $g$.
It turns out that for any $m\in \IZ^+$, the lightest Kaluza-Klein states 
as $\theta\to 0$ are those with $(n_1,n_2)= (f_k, f_{k+1}+m f_{k})$
where $\lbrace f_k \rbrace$ are the same Fibonacci numbers as before.
Indeed, it is easy to verify that these states continue to 
satisfy Eq.~(\ref{numbtheory}) with $\gamma=0$ and $A=1/\sqrt{5}$ for all $m$.
However, while the value of $A$ does not change as a function of $m$,
we can make $|\tau|$ arbitrarily large simply by choosing $m$
arbitrarily large!
Thus, the mass gap ratio $2A|\tau|$ can be made to increase
without bound.
In other words, the phenomenological effects of taking $\theta\to 0$ 
become more and more significant 
as $m\to \infty$.

It should be noted that while the mass gap ratio $\mu'/\mu$ in
Eq.~(\ref{gapratio}) increases
with increasing $m$, the mass gap $\mu$ at $\theta=\pi/2$ 
actually decreases.  This is required by the fact that
$A$ itself (and hence $\mu'$ itself) is bounded from above.
However, the fact that the mass gap {\it ratio}\/ can become arbitrarily
large illustrates the fact  that shape
parameters can have a profound effect in 
altering the properties of the Kaluza-Klein spectrum.
Moreover, while the original 
mass gap at $\theta=\pi/2$ depends strongly on $|\tau|$, we have seen that
the mass gap as $\theta\to 0$ becomes {\it independent}\/ of $|\tau|$.  
We shall see below that this has important phenomenological implications.

We also stress that none of these conclusions rely on choosing
values of $|\tau|$ which are related to the golden 
mean $g=\half (1+\sqrt{5})$.
Indeed, for {\it any}\/ algebraic irrational number $\xi$, there
is always a corresponding non-zero value $A_\xi$. 
We can then always find a related set of irrational numbers $|\tau|$ in the same
equivalence class as $\xi$ such that $2A_\xi |\tau|\to \infty$.

\section{Discussion and phenomenological implications}

In many situations, these observations 
can be exploited in order to dramatically
weaken the experimental bounds on scenarios involving large extra dimensions.

To see this, let us consider the scenario of Ref.~\cite{ADD}
in which large extra dimensions felt only by gravity are responsible
for lowering the fundamental (higher-dimensional) Planck scale into the
TeV range.  In this scenario,
the ratio between the four-dimensional
and higher-dimensional
Planck scales is set purely by the compactification volume $V$;
the shape moduli are irrelevant in this regard.  (This is evident from
the usual Gauss-law arguments~\cite{ADD};  essentially the higher-dimensional
limit is achieved by considering length scales so small that the precise
shape of the compactification manifold becomes irrelevant.)
We are therefore free to choose our shape moduli so as to avoid laboratory,
astrophysical, and cosmological constraints.

It is precisely here that our observations come into play.
Let us first consider the experimental bounds on extra
dimensions that would normally apply in the case with $\theta=\pi/2$.
In this case, we know from Eq.~(\ref{KKmassestwo})
that the lightest Kaluza-Klein states have masses
\beq
    \left( {V\over 4\pi^2 }\right) M_{n_1,n_2}^2  ~=~ {1\over |\tau|}~
\label{usuallowerbound}
\eeq
where we have assumed $|\tau|\equiv R_2/R_1 \geq 1$
without loss of generality.
However, no Kaluza-Klein states have ever been detected experimentally;
thus the mass of the lightest Kaluza-Klein states must exceed
some experimental limit $M_{\rm expt}$.
For example,
in the case of extra dimensions felt only by gravity, 
the current bound~\cite{adelberger} is given by $M_{\rm expt}\sim ({\rm mm})^{-1}$.
Thus, demanding $M_{n_1,n_2}\geq M_{\rm expt}$ in 
Eq.~(\ref{usuallowerbound}),
we find
\beq
      V|\tau| ~\leq~ 4\pi^2 (M_{\rm expt})^{-2}~.
\label{constraint1}
\eeq
Note that since $V\equiv 4\pi^2 R_1 R_2$ and $|\tau|\equiv R_2/R_1 >1$,
this constraint equivalently takes the form
\beq
      R_2^2 ~\leq~  (M_{\rm expt})^{-2}~.
\label{constraint2}
\eeq
Thus, we see that it is not the {\it volume}\/
that is bounded when $\theta=\pi/2$  ---  strictly speaking,
what is actually bounded is the size of the largest single radius.

By contrast, let us now consider the same situation as $\theta\to 0$.
In this case, we have already seen that Eq.~(\ref{usuallowerbound})
is replaced by Eq.~(\ref{lowerbound}) where $A >0$.
Thus, again imposing the experimental constraint that the lightest
Kaluza-Klein states have masses exceeding $M_{\rm expt}$, 
we find that Eq.~(\ref{constraint1}) 
is replaced by
\beq
      V ~\leq~  8\pi^2\,A\, (M_{\rm expt})^{-2}~.
\label{constraint3}
\eeq
Equivalently, multiplying both sides of Eq.~(\ref{constraint3}) by $|\tau|$,
we see that Eq.~(\ref{constraint2}) is replaced by
\beq
        R_2^2 ~\leq~ { 2A|\tau|\over \sin\theta} \, (M_{\rm expt})^{-2}~.
\label{constraint4}
\eeq
However, we have already demonstrated that it is possible to choose $|\tau|$ such
that the product $A |\tau|$ becomes arbitrarily large.
Moreover, in the limit we are considering, $\sin\theta\to 0$.
Thus, in such cases, the experimental bounds when 
$\theta\to 0$ become {\it infinitely weaker}\/ than  
they are when $\theta=\pi/2$.

To phrase this result somewhat differently,  
note that whereas Eq.~(\ref{constraint1}) is really a bound on 
the single largest radius,
Eq.~(\ref{constraint3}) is actually a bound on the compactification {\it volume}\/.
As such, it is completely insensitive to the size of the largest radius!
As long as $|\tau|\equiv R_2/R_1$ is chosen within a fixed 
equivalence class of algebraic irrational numbers (so that the 
corresponding value of $A$ remains fixed), 
we can choose $R_2$ as large as we wish without running afoul of 
experimental constraints! 
The root of this result, of course, is our critical observation 
in Eq.~(\ref{lowerbound})
that the Kaluza-Klein mass gap 
becomes independent of $|\tau|$ in the $\theta\to 0$ limit. 

This result is striking.
After all, even in the usual case with $\theta=\pi/2$, we know that we retain
the freedom to change the ratio $|\tau|\equiv  R_2/R_1$ while keeping the volume
fixed.  However, this freedom can usually be exploited {\it only up to a point}\/:  
no single radius can exceed the size $(M_{\rm expt})^{-1}$ set by the experimental
constraints.
By contrast, as $\theta\to 0$, we see that we can adjust this ratio $|\tau|$ at will, 
making either radius as large as we wish while holding the volume fixed.  
Indeed, as long as $|\tau|$ is chosen from within the same equivalence class 
of algebraic irrational numbers, there is no experimental limit on the size to which the 
single largest dimension can grow.
Thus, in this sense, a large extra dimension can truly be rendered ``invisible'',
even when the compactification volume is held fixed.

This result implies that we must be extremely careful when interpreting the
results of precision tests of non-Newtonian gravity~\cite{adelberger}.
Indeed, we now see that such tests need not be interpreted as placing
limits merely on the size of the largest extra dimension;
in some cases they instead place limits directly on the compactification
volume, leaving the compactification radii completely unconstrained.
This observation also applies to astrophysical bounds that come 
from Kaluza-Klein graviton production (\eg, in supernovas),
and likewise to bounds that may come from 
Kaluza-Klein neutrinos~\cite{neutrinos},
axions~\cite{axions}, and other bulk fields.

One might 
worry that if this scenario is embedded into string theory,
there might exist winding-mode states which become light when the 
larger radius becomes infinitely large and 
the smaller radius becomes correspondingly small. 
This would certainly be the case if the Kaluza-Klein
masses were to become heavier than the fundamental string scale.  However, as $\theta\to 0$,
we have seen that the masses of the lightest Kaluza-Klein states do not 
grow arbitrarily large;  instead, these masses are bounded from above
because $A$ itself is bounded from above.
Thus, as long as the Kaluza-Klein states remain light compared to the fundamental string
scale, the winding-mode states remain correspondingly heavy.

Another potential worry concerns the existence of toroidal
modular symmetries.
Through modular transformations, it is always possible to redefine
the values of $R_1$, $R_2$, and $\theta$ without 
changing any of the underlying physics.
However, such modular transformations
necessarily leave the Kaluza-Klein spectrum invariant.
Thus, any effects which actually modify the Kaluza-Klein 
spectrum (such as the shape effects we have been studying)
must represent physical effects which go beyond mere modular transformations.
Indeed, the critical distinction 
between rational and irrational values of $|\tau|$ 
which we have observed as $\theta\to 0$
is not something that an $SL(2,\IZ)$  modular transformation  
(with its integer coefficients) can eliminate.
This issue will be discussed further in Ref.~\cite{longpaper}.

Clearly, the scenario we have outlined in this paper rests upon 
the unique properties exhibited by the Kaluza-Klein spectrum that emerge 
as $\theta\to 0$ when $|\tau|$ is chosen to be an algebraic irrational number. 
While it may seem unnatural to take such small values 
of $\theta$, we stress that they are part of the allowed compactification moduli space
as long as $\theta>0$. 
Likewise, it may seem fine-tuned to take $|\tau|\equiv R_2/R_1$ irrational.
However, given that $(R_1,R_2)$ are {\it a priori}\/ unconstrained,
all values are equally likely, and indeed it is the {\it rational}\/
values which represent fine-tuning.

In fact, given the modular symmetries of the torus, such moduli may even
be {\it preferred}\/.  As we have mentioned above, the Kaluza-Klein spectrum 
in Eq.~(\ref{KKmassestwo}) is invariant under the modular 
transformations $\tau\to \tau+1$
[under which $(n_1,n_2)\to (n_1,n_2-n_1)$] and $\tau\to -1/\tau$
[under which $(n_1,n_2)\to (-n_2,n_1)$].
Since these modular transformations do not change the underlying torus,
they should be a symmetry of any
dynamical effective potential $V_{\rm eff}(\tau)$
which eventually stabilizes the moduli, with
the extrema of the potential corresponding
to fixed points under the modular transformations.
(An example of this in the case of 
Casimir energies can be found in Ref.~\cite{poppitz}.)
For $\tau_2>0$, it is well known that there are only two distinct
fixed points:  $\tau=i$ and $\tau= e^{\pi i/3}$.  
However, if we permit ourselves to consider the $\tau_2\to 0$ limit,
we find that there are a series of additional fixed points
with 
\beq
     |\tau| ~=~\tau_1~=~  \half \left( p \pm \sqrt{p^2 -4}\right)~,~~~ p\in\IZ \geq 2~.
\eeq
Indeed, for these points the modular transformation $\tau\to -1/\tau$
produces $\tau_1\to \tau_1 +p$, which is identified with $\tau_1$
under the torus symmetries.  
Note that for $p\geq 2$, these values are all algebraic irrational numbers, as desired.
Moreover, this series of fixed points
includes the value $|\tau|=\half (3+\sqrt{5})$ which, as we have already seen,
corresponds to the maximum possible value of $A$.
Thus, even though such points with $\tau_2= 0$
are at the ``edge'' of the allowed moduli space (and strictly speaking
are not even within the fundamental domain of the modular group), we can imagine that
the dynamics might cause the shape moduli to {\it approach}\/ these 
fixed points given appropriate initial conditions. 
As such, these limiting cases could emerge as the result of 
non-perturbative string dynamics or via cosmological evolution.

In summary, then, we have shown that the shape moduli associated with
large-radius compactifications can have a significant effect on the corresponding
Kaluza-Klein spectrum and in turn on the resulting low-energy phenomenology.
We investigated these ideas in the context of flat, two-dimensional toroidal 
compactifications, but similar effects are also likely to arise in more 
complicated higher-dimensional compactifications on more exotic manifolds~\cite{Kaloper},
or even in ``warped'' compactifications~\cite{RS}.
Indeed, we have seen that in certain limiting cases, it is possible
to make large extra dimensions essentially ``invisible'' with respect to experimental
and observational constraints on light Kaluza-Klein states.
The incorporation of shape moduli can therefore be used to significantly
widen the allowed parameter
space of higher-dimensional theories beyond 
what has previously been considered.

\section*{Acknowledgments}

This work was supported in part by the National Science Foundation
under Grant PHY-0071054, and by a Research Innovation Award from 
Research Corporation. 
I would like to thank C.~Cs\'aki, J.~Erlich, A.~Peet, E.~Poppitz, S.~Sethi,
H.~Tye, and A.~Vasiu for discussions. 
I would especially like to thank T.~Gherghetta, A.~Mafi, and W.~McCallum 
for comments and feedback concerning the ideas put forth in this paper. 
I also wish to acknowledge the hospitality of the Aspen Center for 
Physics where this work was performed.




\begin{references}


\bibitem{ADD}   N.~Arkani-Hamed, S.~Dimopoulos and G.~Dvali,
           Phys.\ Lett.\ B {\bf 429} (1998) 263
          [hep-ph/9803315];
            I.~Antoniadis {\it et al.}, 
             Phys.\ Lett.\ B {\bf 436} (1998) 257
              [hep-ph/9804398].

\bibitem{DDG}
         K.R.~Dienes, E.~Dudas and T.~Gherghetta,
           Phys.\ Lett.\ B {\bf 436} (1998) 55
          [hep-ph/9803466];
           Nucl.\ Phys.\ B {\bf 537} (1999) 47
               [hep-ph/9806292];
           hep-ph/9807522.

\bibitem{string}
          E.~Witten, Nucl.\ Phys.\ B {\bf 471} (1996) 135
             [hep-th/9602070];
          J.D.~Lykken, Phys.\ Rev.\ D {\bf 54} (1996) 3693
            [hep-th/9603133];
          G.~Shiu and S.-H.H.~Tye, Phys.\ Rev.\ D {\bf 58} (1998) 106007
             [hep-th/9805157];
          Z.~Kakushadze and S.-H.H.~Tye,
              Nucl.\ Phys.\ B {\bf 548} (1999) 180
             [hep-th/9809147].

\bibitem{kehagias}
           A.~Kehagias and K.~Sfetsos,
          Phys.\ Lett.\ B {\bf 472} (2000) 39
           [hep-ph/9905417];
         E.G.~Floratos and G.K.~Leontaris,
           Phys.\ Lett.\ B {\bf 465} (1999) 95
            [hep-ph/9906238].

\bibitem{textbook}
         See, {\it e.g.}\/, 
         G.H.~Hardy and E.M.~Wright, {\it An Introduction
          to the Theory of Numbers, 4th Edition}\/  
       (Oxford University Press, 1959);
       M.~Hindry and J.H.~Silverman,
       {\it Diophantine Geometry:  An Introduction}\/,
       Graduate Texts in Mathematics \#201
     (Springer-Verlag, 2000).

\bibitem{Roth}
         K.F.~Roth, {\it Mathematika}\/ {\bf 2} (1955)~1.

\bibitem{adelberger}
        See, \eg, C.D.~Hoyle {\it et al.}, 
          Phys.\ Rev.\ Lett.\  {\bf 86} (2001) 1418
          [hep-ph/0011014].


\bibitem{neutrinos}
        K.R.~Dienes, E.~Dudas and T.~Gherghetta,
          Nucl.\ Phys.\ B {\bf 557} (1999) 25
          [hep-ph/9811428];
          N.~Arkani-Hamed {\it et al.}, 
             hep-ph/9811448.


\bibitem{axions}
   S.~Chang, S.~Tazawa and M.~Yamaguchi,
     Phys.\ Rev.\ D {\bf 61} (2000) 084005
       [hep-ph/9908515];
   K.R.~Dienes, E.~Dudas and T.~Gherghetta,
        Phys.\ Rev.\ D {\bf 62} (2000) 105023
        [hep-ph/9912455];
       L.~Di Lella {\it et al.}\/,
       Phys.\ Rev.\ D {\bf 62} (2000) 125011
       [hep-ph/0006327].


\bibitem{longpaper}
      K.R.~Dienes and A.~Mafi, {\it Compactification on
        Manifolds with Non-Trivial Shape Moduli}\/,
        to appear.


\bibitem{poppitz}
        E.~Pont\'on and E.~Poppitz, JHEP {\bf 0106} (2001) 019
               [hep-ph/0105021].

\bibitem{Kaloper}
        See, {\it e.g.}\/, N.~Kaloper {\it et al.}, 
          Phys.\ Rev.\ Lett.\  {\bf 85} (2000) 928
        [hep-ph/0002001].
 

\bibitem{RS}
        L.~Randall and R.~Sundrum,
         Phys.\ Rev.\ Lett.\  {\bf 83} (1999) 3370
          [hep-ph/9905221];
         Phys.\ Rev.\ Lett.\  {\bf 83} (1999) 4690
         [hep-th/9906064].




\end{references}
\end{document}